\documentclass[pra,aps,twocolumn,amsmath,amssymb,showkeys,showpacs]{revtex4}
\usepackage{graphicx}
\newcommand{\tr}{\mathrm{Tr}}
\newcommand{\hm}{\hat{H}}

\newcommand{\vm}{\hat{V}}
\newcommand{\xm}{\hat{X}}

\newcommand{\ip}{{\hat{\Pi}}}
\newcommand{\ssf}{{\hat{S}}}

\newcommand{\pen}{\openone}

\newcommand{\bro}{\hat{\rho}}

\newcommand{\ca}{\mathcal{C}}
\newcommand{\tca}{\mathcal{R}}

\newcommand{\iu}{{\mathtt{i}}}

\begin{document}
\clearpage
\preprint{}

\title{Formulation of Leggett--Garg inequalities in terms of $q$-entropies}
\author{Alexey E. Rastegin}
\email{rast@api.isu.ru; alexrastegin@mail.ru}
\affiliation{Department of Theoretical Physics, Irkutsk State University,
Gagarin Bv. 20, Irkutsk 664003, Russia}

\begin{abstract}
As is well known, the macroscopic realism and the noninvasive
measurability together lead to Leggett--Garg inequalities violated
by quantum mechanics. We consider tests of the Leggett--Garg type
with use of the $q$-entropies. For all $q\geq1$, quantum mechanics
predicts violations of an entire family of $q$-entropic
inequalities of the Leggett--Garg type. Violations are exemplified
with a quantum spin-$s$ system. In general, entropic Leggett--Garg
inequalities give only necessary conditions that some
probabilistic model is compatible with the macrorealism in the
broader sense. The presented $q$-entropic inequalities allow to
widen a class of situations, in which an incompatibility with the
macrorealism can be tested. In the considered example, both the
strength and range of violations are somehow improved by varying
$q$. We also examine $q$-entropic inequalities of the
Leggett--Garg type in the case of detection inefficiencies, when
the no-click event may occur in each observation. With the use of
the $q$-entropic inequalities, the required amount of efficiency
may be reduced.
\end{abstract}

\maketitle

\pagenumbering{arabic}
\setcounter{page}{1}

\section{Introduction}\label{sec1}

Physicists know a few key advances that emphasize distinctions of
the quantum world from the classical one. The uncertainty
principle was a primary among them \cite{wh27}. The Bell theorem
\cite{bell64} is a next profound insight into the subject. It is
closely related to the Einstein--Podolsky--Rosen question
\cite{epr35} and later reformulation by Bohm \cite{bohm51}.
Studies of foundations of quantum theory are now connected with a
progress in quantum information processing \cite{beh2013}.
Violations of Bell inequalities reveal non-classical features of
correlations between spatially-separated quantum systems
\cite{mermin93}. The Clauser--Horne--Shimony--Holt (CHSH) scenario
\cite{chsh69} is the first setup tested in experiments
\cite{agr82,adr82}. Violations of the CHSH inequality imply that
predictions of quantum theory are not compatible with the local
realism \cite{az99}. The Klyachko--Can--Binicio\v{g}lu--Shumovsky
(KCBS) scenario \cite{kly08} pertains to the measurement
statistics of a single spin-$1$ system. Since made experiments
gave expected results, non-local hidden-variable theories become
the subject of researches \cite{ajl03}.

Leggett--Garg inequalities \cite{lg85} form one of directions
inspired by the Bell theorem. These inequalities are based on the
following two concepts often called the macrorealism in the
broader sense. First, we assume that physical properties of a
macroscopic object preexist irrespectively to the act of
observation. Second, measurements are noninvasive in the sense
that the measurement of an observable at any instant of time does
not alert its subsequent evolution. Consequences of the
assumptions were originally examined by Leggett and Garg
\cite{lg85}. They are commonly known as Leggett--Garg inequalities
\cite{aln13}. It turns out that predictions of quantum mechanics
lead to violations of these inequalities. Leggett--Garg
inequalities are now the subject of active experimental
\cite{dbhj11,arm11,kat13} and theoretical investigations
\cite{b09,ahw10,wm12,arav13,dk14}. In practice, decoherence is one
of crucial problems. Experimental violations of the Leggett--Garg
inequalities under decoherence are considered in Refs.
\cite{pal10,xu11}. Interesting physical proposals are discussed in
Refs. \cite{wj08,eln12,ghs13}.

Entropic approach to formulating the Bell theorem was proposed in
Ref. \cite{BC88} and later studied in Refs.
\cite{cerf97,rchtf12,krk12}. In particular, entropic inequalities
of the Bell type were derived for the KCBS scenario
\cite{rchtf12,krk12}. Entropic formulations are very useful due to
the following. First, they can deal with any finite number of
outcomes. Second, entropic approach allows to address more
realistic cases with detection inefficiencies \cite{rchtf12}.
Additional possibilities to analyze non-locality or contextuality
or probabilistic models are provided by use of the $q$-entropies
\cite{rastqic14}. Using the $q$-entropic inequalities, we can
widen a class of probability distributions, for which the
non-locality or contextuality are testable. It is an alternative
to the approach with adding some shared randomness \cite{rch13}.
Further, the $q$-entropic inequalities are expedient in analyzing
cases with detection inefficiencies \cite{rastqic14}.

Leggett--Garg tests probe the correlations of a single system
measured at different times. It is appealing to study restrictions
of the Leggett--Garg type within an entropic approach. Using
standard entropic functions, such an analysis has been carried out
by the writers of Ref. \cite{uksr12}. In the present paper, we aim
to study restrictions of the Leggett--Garg type with formulating
them in terms of the Tsallis $q$-entropies. Our paper is organized
as follows. In Section \ref{sec2}, we recall basic properties of
the $q$-entropies. Leggett--Garg inequalities in terms of the
$q$-entropies are derived in Section \ref{sec3}. We also consider
a formulation of entropic Leggett--Garg inequalities in the case
of detection inefficiencies. In Section \ref{sec4}, violations of
the derived Leggett--Garg inequalities are exemplified with a
quantum spin-$s$ system. We also discuss trade-offs between
violations of the $q$-entropic inequalities and the required
efficiency of detectors. In Section \ref{sec5}, we conclude the
paper with a summary of results.

\section{Tsallis $q$-entropies and their properties}\label{sec2}

In this section, we recall some preliminary material on the
Tsallis $q$-entropies and their properties. Let $X$ be discrete
random variable taking values according to the probability
distribution $\bigl\{p(x):{\>}x\in\Omega_{X}\bigr\}$. The Tsallis
entropy of degree $q>0\neq1$ is defined by \cite{tsallis}
\begin{equation}
H_{q}(X):=\frac{1}{1-q}{\,}\left({\sum_{x\in\Omega_{X}}p(x)^{q}} - 1 \right)
{\,}. \label{tsaent}
\end{equation}
With slightly other factor, this entropic function was proposed by
Havrda and Charv\'{a}t \cite{havrda}. Let $Y$ be another variable
taking values with the probability distribution
$\bigl\{p(y):{\>}y\in\Omega_{Y}\bigr\}$. The joint $q$-entropy
$H_{q}(X,Y)$ is defined similarly to Eq. (\ref{tsaent}), but with
joint probabilities $p(x,y)$. We rewrite the entropy
(\ref{tsaent}) in the form
\begin{align}
H_{q}(X)&=-\sum_{x\in\Omega_{X}}{p(x)^{q}\ln_{q}p(x)}
\nonumber\\
&=\sum_{x\in\Omega_{X}}{p(x){\>}{\ln_{q}}{\left(\frac{1}{p(x)}\right)}}
{\,} . \label{tsaln}
\end{align}
Here, the $q$-logarithm
$\ln_{q}(\xi)=\bigl(\xi^{1-{q}}-1\bigr)/(1-{q})$ is defined for
$q>0\not=1$ and $\xi>0$. In the limit $q\to1$, we obtain
$\ln_{q}(\xi)\to\ln\xi$ and the Shannon entropy
\begin{equation}
H_{1}(X)=-\sum\nolimits_{x\in\Omega_{X}}{p(x){\,}\ln{p(x)}}
\ . \label{shaln}
\end{equation}
For brevity, we will usually omit the range of summation. The
entropy (\ref{tsaent}) is widely used in many disciplines
\cite{gmt}. The R\'{e}nyi entropies \cite{renyi61} form another
especially important family of generalized entropies. Applications
of these entropies and their quantum counterparts are considered
in the book \cite{bengtsson}.

To consider cases with detector inefficiencies, the following
question will rise \cite{rchtf12}. In real experiments, we do not
deal immediately with original distributions of the form
$\bigl\{p(x)\bigr\}$. Such distributions will somehow be altered
due to detector inefficiencies. To the given $\eta\in[0;1]$ and
probability distribution $\bigl\{p(x):{\>}x\in\Omega_{X}\bigr\}$,
we assign another probability distribution
\begin{equation}
\bigl\{\eta{p}(x):{\>}x\in\Omega_{X}\bigr\}\cup\{1-\eta\}
\ . \label{peta}
\end{equation}
This probability distribution corresponds to some ``altered''
random variable $X_{\eta}$. For all $q>0$, the entropy
$H_{q}(X_{\eta})$ can be expressed as \cite{rastqic14}
\begin{equation}
H_{q}(X_{\eta})=\eta^{q}H_{q}(X)+h_{q}(\eta)
\ . \label{rlem0}
\end{equation}
As usually, the binary $q$-entropy reads
\begin{equation}
h_{q}(\eta):=-{\,}\eta^{q}\ln_{q}(\eta)-(1-\eta)^{q}\ln_{q}(1-\eta)
\ . \label{bend}
\end{equation}
From three probability distributions, we can built another
probability distribution
\begin{align}
\{p_{\eta\eta}\}&:=\bigl\{\eta^{2}p(x)\bigr\}\cup\bigl\{\eta(1-\eta)p(y)\bigr\}
\nonumber\\
&\cup\bigl\{\eta(1-\eta)p(z)\bigr\}\cup\bigl\{(1-\eta)^{2}\bigr\}
\ . \label{prete}
\end{align}
It is assigned to some random variable $X_{\eta\eta}$. For all
$q>0$, we then have \cite{rastqic14}
\begin{align}
H_{q}(X_{\eta\eta})&=
\eta^{2q}H_{q}(X)+\eta^{q}(1-\eta)^{q}\bigl(H_{q}(Y)+H_{q}(Z)\bigr)
\nonumber\\
&+\bigl(\eta^{q}+(1-\eta)^{q}+1\bigr)h_{q}(\eta)
\ . \label{rlem00}
\end{align}
We will use the results (\ref{rlem0}) and (\ref{rlem00}) for
studying entropic Leggett--Garg inequalities in the case of
detection inefficiencies.

Like the Braunstein--Caves inequality \cite{BC88}, entropic
Leggett--Garg inequalities are formulated in terms of the
conditional entropy \cite{uksr12}. The entropy of $X$ conditional
on knowing $Y$ is defined as \cite{CT91}
\begin{align}
H_{1}(X|Y)&:=\sum\nolimits_{y}{p(y){\,}H_{1}(X|y)}
\nonumber\\
&=-\sum\nolimits_{x}\sum\nolimits_{y}{p(x,y){\,}\ln{p}(x|y)}
\ . \label{cshen}
\end{align}
Here, we take $H_{1}(X|y):=-\sum_{x}{p(x|y)\ln{p}(x|y)}$ and
$p(x|y)=p(x,y){\,}p(y)^{-1}$ according to Bayes's rule. The
quantity (\ref{cshen}) is the standard conditional entropy. For
partitions on quantum logic, the standard conditional entropies
were studied in Ref. \cite{zhma}. Further development with the use
of the R\'{e}nyi and Tsallis entropies was reported in Ref.
\cite{rastctp}.

We recall the $q$-entropic extension of Eq. (\ref{cshen}).
Introducing the particular functional
\begin{equation}
H_{q}(X|y):=-\sum\nolimits_{x}{p(x|y)^{q}\ln_{q}p(x|y)}
\ , \label{pcen}
\end{equation}
we define the conditional $q$-entropy as \cite{sf06,rastkyb}
\begin{equation}
H_{q}(X|Y):=\sum\nolimits_{y}{p(y)^{q}{\,}H_{q}(X|y)}
\ . \label{qshen}
\end{equation}
Taking the limit $q\to1$, this definition leads to Eq.
(\ref{cshen}). Below, we will mainly use the following properties.
For all $q>0$, the entropy (\ref{qshen}) satisfies
\begin{align}
H_{q}(X,Y)&=H_{q}(Y|X)+H_{q}(X)
\nonumber\\
&=H_{q}(X|Y)+H_{q}(Y)
\ . \label{chrl}
\end{align}
It is referred to as the chain rule for the conditional
$q$-entropy \cite{sf06}. By theorem 2.4 of Ref. \cite{sf06},
we have the chain rule with a finite number of random variables:
\begin{equation}
H_{q}(X_{1},X_{2},\ldots,X_{n})=
\sum_{j=1}^{n}H_{q}(X_{j}|X_{j-1},\ldots,X_{1})
\ . \label{ehrl}
\end{equation}
For real $q\geq1$ and integer $n\geq1$, the conditional
$q$-entropy also satisfies \cite{sf06,rastqic14}
\begin{equation}
H_{q}(X|Y_{1},\ldots,Y_{n-1},Y_{n})
\leq{H}_{q}(X|Y_{1},\ldots,Y_{n-1})
\ . \label{rlem1}
\end{equation}
Due to Eq. (\ref{rlem1}), conditioning on more can only reduce the
$q$-entropy of degree $q\geq1$. In Ref. \cite{rastqic14}, we
examined formulation of Bell's theorem in terms of the
$q$-entropies. In a similar manner, we will study the macrorealism
in the broader sense with use of the conditional $q$-entropies.
Deriving $q$-entropic forms of Leggett--Garg inequalities will be
based on the properties listed above.

\section{Entropic Leggett--Garg inequalities in terms of $q$-entropies}\label{sec3}

We begin with discussion of basic points involved in the
macrorealistic picture. Leggett--Garg inequalities are based on
the following two assumptions known as the macroscopic realism and
the noninvasive measurability at the macroscopic level
\cite{aln13}. We consider a macrorealistic system, in which
$X(t_{j})$ is a dynamical variable at the time moment $t_{j}$.
Formally, the macroscopic realism {\it per se} implies that
outcomes $x_{j}$ of the variables $X(t_{j})$ at all instants of
time preexist irrespective of their measurements. The noninvasive
measurability means that the act of measuring $X(t_{j})$ at an
earlier time $t_{j}$ does not affect its subsequent value at a
later time $t_{k}>t_{j}$. These assumptions lead to the following
conclusion. For each particular choice of time instants, the
statistics of outcomes is described by a joint probability
distribution $p(x_{1},x_{2},\ldots,x_{n})$. The joint
probabilities are expressed as a convex combination of the form
\cite{kb08,kb12}
\begin{align}
&p(x_{1},x_{2},\ldots,x_{n})=
\nonumber\\
&\sum\nolimits_{\lambda}
{\varrho(\lambda){\,}P(x_{1}|\lambda){\,}P(x_{2}|\lambda)\cdots{P}(x_{n}|\lambda)}
\ . \label{x1x2}
\end{align}
Here, the product of conditional probabilities $P(x_{j}|\lambda)$
is averaged over a hidden-variable probability distribution. Of
course, in any macrorealistic model the probabilities
$P(x_{j}|\lambda)\geq0$ should obey
\begin{equation}
\sum\nolimits_{x_{j}}P(x_{j}|\lambda)=1
\ . \label{socn}
\end{equation}
Further, unknown hidden-variable probabilities
$\varrho(\lambda)\geq0$ should satisfy
$\sum_{\lambda}{\varrho(\lambda)}=1$. By a structure, the
$n$-variable distribution (\ref{x1x2}) will marginalize to
particular distributions with lesser number of variables. This is
a consistency condition for macrorealistic models.

Like probabilistic model of the local realism and
noncontextuality, the existence of joint probability distributions
of the form (\ref{x1x2}) does result in certain inequalities
between conditional entropies. Entropic inequalities of Ref.
\cite{uksr12} were derived similarly to the treatment given by
Braunstein and Caves \cite{BC88}. For the CHSH and KCBS scenarios,
the $q$-entropic inequalities were formulated in Ref.
\cite{rastqic14}. Let us apply these ideas to macrorealistic
models. We will use $X_{j}$ as shortening for $X(t_{j})$. For
brevity, we consider the case $n=3$ with the variables $X_{1}$,
$X_{2}$, $X_{3}$. For $q\geq1$, one gets
\begin{align}
&H_{q}(X_{1},X_{3})\leq{H}_{q}(X_{1},X_{2},X_{3})
\nonumber\\
&=H_{q}(X_{1})+H_{q}(X_{2}|X_{1})+H_{q}(X_{3}|X_{2},X_{1})
\nonumber\\
&\leq{H}_{q}(X_{1})+H_{q}(X_{2}|X_{1})+H_{q}(X_{3}|X_{2})
\ . \label{cmb01}
\end{align}
Here, we used the chain rule (\ref{ehrl}) and suitable relations
of the form (\ref{rlem1}). Subtracting $H_{q}(X_{1})$ and using
the chain rule again, we obtain the entropic inequality
\begin{equation}
H_{q}(X_{3}|X_{1})\leq{H}_{q}(X_{3}|X_{2})+H_{q}(X_{2}|X_{1})
\ , \label{cmb11}
\end{equation}
which holds for $q\geq1$. For $q=1$, this formula is reduced to
the Shannon-entropy inequality given in Ref. \cite{uksr12}. By a
parallel argument, for real $q\geq1$ and integer $n\geq3$ we
obtain
\begin{equation}
H_{q}(X_{n}|X_{1})-\sum_{n\geq{j}\geq2}H_{q}(X_{j}|X_{j-1})=:\ca_{q}\leq0
\ . \label{cmb12}
\end{equation}
We introduce here the characteristic quantity $\ca_{q}$. In the
next section, we will exemplify that quantum mechanics sometimes
leads to violations of Eq. (\ref{cmb12}). Positive values of $\ca_{q}$
then characterize an amount with which entropic Leggett--Garg
inequalities are violated. In the case $n=4$, the relation
(\ref{cmb12}) is formally similar to the $q$-entropic version of
the Braunstein--Caves inequality. For $n=5$, the result
(\ref{cmb12}) mathematically coincides with the $q$-entropic
inequalities holding for non-contextual models in the KCBS
scenario. Such $q$-entropic inequalities for both the CHSH and
KCBS scenarios were examined in Ref. \cite{rastqic14}.

Real measurement devices are inevitably exposed to noise. Entropic
approach allows to take into account such a feature. The
Shannon-entropy formulation of Bell's theorem with detection
inefficiencies was considered in Ref. \cite{rchtf12}. In Ref.
\cite{rastqic14}, we extended this treatment to $q$-entropic
inequalities. It is relevant to address the question of detection
inefficiencies also for entropic Leggett--Garg inequalities. For
these purposes, we adopt one of the inefficiency models considered
in Ref. \cite{rchtf12}. Let us assume that the no-click event can
occur in each act of observation irrespectively to other
observations. We also assume that detectors are all of efficiency
$\eta\in[0;1]$. For a pair of outcomes of the dynamical quantities
$X$ and $Y$, we have probabilities
\begin{align}
p^{(\eta\eta)}(x,y)&=\eta^{2}p(x,y)
\ , \label{pretxx}\\
p^{(\eta\eta)}(x,\varnothing)&=\eta(1-\eta)p(x)
\ , \label{pretx1}\\
p^{(\eta\eta)}(\varnothing,y)&=\eta(1-\eta)p(y)
\ , \label{pretx0}\\
p^{(\eta\eta)}(\varnothing,\varnothing)&=(1-\eta)^{2}
\ . \label{pret00}
\end{align}
Here, the no-click event is denoted by ``$\varnothing$''. The
two-variable probability distribution
(\ref{pretxx})--(\ref{pret00}) marginalizes to the
single-observable distributions of the form
\begin{align}
p^{(\eta)}(x)&=\eta{p}(x)
\ , \label{pretx}\\
p^{(\eta)}(\varnothing)&=1-\eta
\ . \label{prety}
\end{align}
Let us rewrite Eq. (\ref{cmb12}) without conditional entropies.
Using Eq. (\ref{chrl}), we finally get the theoretical result
\begin{align}
{\ }-\ca_{q}&=\sum_{j=1}^{n-1}H_{q}(X_{j},X_{j+1})
-H_{q}(X_{1},X_{n})
\nonumber\\
& -\sum_{k=2}^{n-1}H_{q}(X_{k})\geq0
{\>}. \label{cmb21}
\end{align}
In Eq. (\ref{cmb21}), all the entropies pertain to the
inefficiency-free case, when $\eta=1$. However, we actually deal
with ``altered'' probability distributions described by the
formulas (\ref{pretxx})--(\ref{pret00}) and
(\ref{pretx})--(\ref{prety}). Using the results (\ref{rlem0}) and
(\ref{rlem00}), we obtain
\begin{align}
H_{q}^{(\eta)}(X_{k})&=\eta^{q}H_{q}(X_{k})+h_{q}(\eta)
\ , \label{clem0b}\\
H_{q}^{(\eta\eta)}(X_{j},X_{j+1})&=\eta^{2q}H_{q}(X_{j},X_{j+1})
\nonumber\\
&+\eta^{q}(1-\eta)^{q}\bigl(H_{q}(X_{j})+H_{q}(X_{j+1})\bigr)
\nonumber\\
&+\bigl(\eta^{q}+(1-\eta)^{q}+1\bigr)h_{q}(\eta)
\ . \label{clem1}
\end{align}
By $H_{q}^{(\eta)}(X_{k})$ and
$H_{q}^{(\eta\eta)}(X_{j},X_{j+1})$, we mean the actual
$q$-entropies calculated with the probability distributions
(\ref{pretx})--(\ref{prety}) and (\ref{pretxx})--(\ref{pret00}).
Instead of the characteristic quantity $\ca_{q}$, we will deal
with
\begin{align}
\ca_{q}^{(\eta\eta)}&:=H_{q}^{(\eta\eta)}(X_{1},X_{n})-\sum_{j=1}^{n-1}H_{q}^{(\eta\eta)}(X_{j},X_{j+1})
\nonumber\\
&+\sum_{k=2}^{n-1}H_{q}^{(\eta)}(X_{k})
\ . \label{emb12}
\end{align}
By calculations, we obtain the relations
\begin{align}
\ca_{q}^{(\eta\eta)}&=\eta^{2q}{\,}\ca_{q}-\Delta_{q}(\eta)
\ , \label{caet}\\
\Delta_{q}(\eta)&=
\eta^{q}\bigl(\eta^{q}+2(1-\eta)^{q}-1\bigr)\sum_{k=2}^{n-1}H_{q}(X_{k})
\nonumber\\
&+(n-2)\bigl(\eta^{q}+(1-\eta)^{q}\bigr)h_{q}(\eta)
\ . \label{etac}
\end{align}
For the case $n=5$, the additional term (\ref{etac}) was studied
in Ref. \cite{rastqic14}. When $q>1$, the factor
$\eta^{q}+2(1-\eta)^{q}-1$ is negative for some values of $\eta$
near $1$ from below. Thus, the first term in the right-hand side
of Eq. (\ref{etac}) can take positive or negative values. The
second term in the right-hand side of Eq. (\ref{etac}) is
certainly positive.

The Leggett--Garg inequality (\ref{cmb12}) implies $\ca_{q}\leq0$.
Using measurement statistics, we have to analyze the quantity
(\ref{caet}). Assume that measurement data have lead to the result
$\ca_{q}^{(\eta\eta)}>0$. In principle, we still cannot claim
$\ca_{q}>0$. We must beforehand confide that the violating term
$\eta^{2q}{\,}\ca_{q}$ is essentially larger than the additional
term (\ref{etac}). For very high values of the efficiency parameter
$\eta$, the term (\ref{etac}) will be small. At the same time,
values of $\Delta_{q}(\eta)$ also depend on the entropic parameter
$q\geq1$. It is instructive to consider these questions within
a concrete example. We will address them in the next section.

\section{Entropic Leggett--Garg inequalities for spin systems}\label{sec4}

In this section, we consider concrete systems, for which the
$q$-entropic Leggett--Garg inequalities are violated. Following
Ref. \cite{uksr12}, we consider a quantum spin-$s$ system.
Initially, it is prepared in the completely mixed state
\begin{equation}
\bro_{*}=\frac{1}{2s+1}\sum_{m=-s}^{+s}
{|s,m\rangle\langle{s},m|}=\frac{1}{2s+1}{\>}\pen
\ . \label{cmst}
\end{equation}
As usually, the states $|s,m\rangle$ are common eigenstates of the
commuting operators
$\ssf^{2}=\ssf_{x}^{2}+\ssf_{y}^{2}+\ssf_{z}^{2}$ and $\ssf_{z}$.
By $\pen$, we denote the identity operator in the $(2s+1)$-dimensional
Hilbert space of the system. We use the standard notation
\begin{align}
\ssf^{2}|s,m\rangle&=s(s+1)\hbar^{2}{\,}|s,m\rangle
\ , \label{sev10}\\
\ssf_{z}|s,m\rangle&=m\hbar{\,}|s,m\rangle
\ . \label{sevl}
\end{align}
Evolution of the system in time is generated by the Hamiltonian
\cite{uksr12}
\begin{equation}
\hm=\omega{\,}\ssf_{y}
\ . \label{hmlt}
\end{equation}
We will consider measurements of $z$-component of the spin. In the
Heisenberg picture, its evolution is represented as
\begin{align}
\xm(t)&=\vm(t)^{\dagger}\ssf_{z}\vm(t)
{\>}, \label{xmvm0}\\
\vm(t)&={\exp}{\bigl(-{\iu}t\hbar^{-1}\hm\bigr)}
{\>}. \label{xmvm}
\end{align}
In this picture, the state (\ref{cmst}) remains unchanged until
the act of observation occurs. For brevity, we introduce rank-one
projectors of the form
\begin{equation}
\ip_{m}(t):=\vm(t)^{\dagger}|s,m\rangle\langle{s},m|{\,}\vm(t)
{\>}. \label{prct}
\end{equation}
With the initial state (\ref{cmst}), the measurement of $\xm(t)$
at the time $t=t_{j}$ leads to the result
$m\in\bigl\{-s,-s+1,\ldots,+s\bigr\}$ with probability
\begin{equation}
p(m)={\tr}{\bigl(\bro_{*}{\,}\ip_{m}(t_{j})\bigr)}=(2s+1)^{-1}
{\>}. \label{prmj}
\end{equation}
Thus, the outcomes are all equiprobable. Due to the projection
postulate, the normalized post-measurement state is written as
\begin{equation}
p(m)^{-1}\ip_{m}(t_{j}){\,}\bro_{*}{\,}\ip_{m}(t_{j})=\ip_{m}(t_{j})
\ . \label{post}
\end{equation}
Hence, the context for next observation is determined. Calculating
the conditional probability of obtaining the outcome $m^{\prime}$
at the next time $t_{k}$, we write
\begin{align}
p(m^{\prime}|m)&={\tr}{\left(\ip_{m}(t_{j}){\,}\ip_{m^{\prime}}(t_{k})\right)}
\nonumber\\
&=\left|\langle{s},m^{\prime}|{\,}{\exp}{\bigl(-{\iu}\theta_{kj}\hbar^{-1}\ssf_{y}\bigr)}|s,m\rangle\right|^{2}
{\,}, \label{cpkj}
\end{align}
where $\theta_{kj}=\omega(t_{k}-t_{j})$. Probabilities
(\ref{cpkj}) are immediately expressed in terms of elements of the
corresponding rotation matrix. These elements also known as the
Wigner small $d$-functions are defined as \cite{bl81}
\begin{equation}
d_{m^{\prime},m}^{(s)}(\theta)=\langle{s},m^{\prime}|{\,}{\exp}{\bigl(-{\iu}\theta\hbar^{-1}\ssf_{y}\bigr)}|s,m\rangle
\ . \label{bldd}
\end{equation}
The elements of rotation matrices are well tabulated. Hence, we
write a useful expression
\begin{equation}
p(m^{\prime}|m)=\bigl|d_{m^{\prime},m}^{(s)}(\theta_{kj})\bigr|^{2}
{\,}. \label{cpkjd}
\end{equation}
Combining Eqs. (\ref{prmj}) and (\ref{cpkjd}) gets the joint
probability distribution for results of two sequential
measurements:
\begin{equation}
p(m,m^{\prime})=p(m){\,}p(m^{\prime}|m)=\frac{1}{2s+1}{\>}\bigl|d_{m^{\prime},m}^{(s)}(\theta_{kj})\bigr|^{2}
{\,}. \label{jpjk}
\end{equation}
Recall that expressions of such a kind have been obtained for the
CHSH scenario with singlet initial state of the spin-$s$ system
\cite{BC88}.

Following Ref. \cite{uksr12}, we now consider equidistant time
intervals. That is, two sequential measurements of $\xm(t)$ are
separated by the interval $t_{j}-t_{j-1}=\Delta{t}$. Let us
define parameters $\Delta\theta=\omega\Delta{t}$ and
$\theta=(n-1)\Delta\theta$, and an auxiliary function
\begin{equation}
F_{q}^{(s)}(\theta):=\frac{(2s+1)^{-q}}{q-1}{\,}\sum_{m=-s}^{+s}
{\biggl(1-\sum_{m^{\prime}=-s}^{+s}{\bigl|d_{m^{\prime},m}^{(s)}(\theta)\bigr|^{2q}}\biggr)}
{\,}. \label{ffundf}
\end{equation}
The quantum-mechanical expressions for the conditional
$q$-entropies are then written as
\begin{align}
H_{q}(X_{j}|X_{j-1})&=F_{q}^{(s)}(\Delta\theta)
\ , \label{xjxj0}\\
H_{q}(X_{n}|X_{1})&=F_{q}^{(s)}(\theta)
\ . \label{xjxj}
\end{align}
It is useful to compare the characteristic quantity (\ref{cmb12})
with the entropic scale $\ln_{q}(2s+1)$. The latter gives the
maximum of the $q$-entropy supported on $2s+1$ points. We will
consider the relative quantity
\begin{equation}
\tca_{q}:=\frac{\ca_{q}}{\ln_{q}(2s+1)}=\frac{F_{q}^{(s)}(\theta)-(n-1){\,}F_{q}^{(s)}(\Delta\theta)}{\ln_{q}(2s+1)}
\ . \label{tcdf}
\end{equation}
By Eq. (\ref{cmb12}), the hypotheses of macroscopic realism and
noninvasive measurability lead to the result $\tca_{q}\leq0$.
Positive values of $\tca_{q}$ will imply that predictions of
quantum mechanics are not compatible with these hypotheses. Since
our aim is to focus on variations of the parameter $q$, we
consider only the simplest choice of $s$ and $n$.

\begin{figure}
\includegraphics[width=8.5cm]{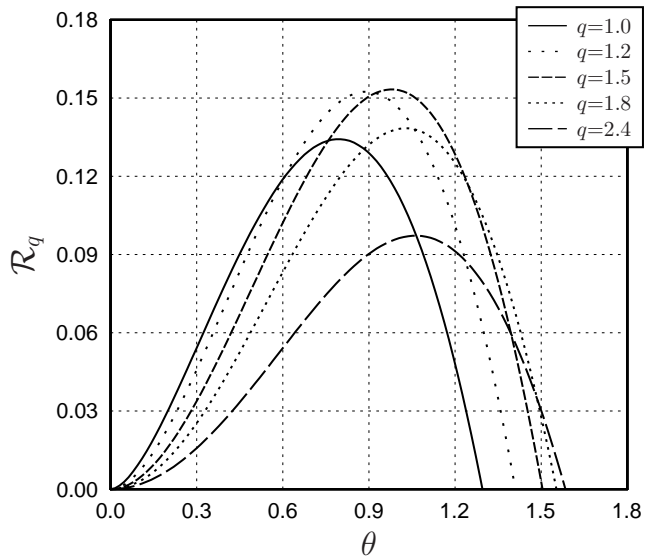}
\caption{\label{fig1}The relative quantity $\tca_{q}$ versus
$\theta$ for $s=1/2$, $n=3$, and five values of $q$, namely
$q=1.0;1.2;1.5;1.8;2.4$. For each value of $q$, only positive
values of $\tca_{q}$ are shown.}
\end{figure}

On Fig. \ref{fig1}, we have shown violations of the $q$-entropic
Leggett--Garg inequalities for the spin $s=1/2$ and the number
$n=3$. The curves are related to the values
$q=1.0;1.2;1.5;1.8;2.4$. The standard choice $q=1$ considered in
Ref. \cite{uksr12} is included for comparison. We see that the curve
maximum goes to larger values of $\theta$ with growth of $q$.
There is some extension of the domain, for which $\tca_{q}>0$. For
$q>2.4$, however, such an extension becomes negligible.
Nevertheless, the curves of Fig. \ref{fig1} clearly show an
utility of the $q$-entropic approach. In this regard, $q$-entropic
inequalities of the Leggett--Garg type are similar to the
$q$-entropic Bell inequalities derived in Ref. \cite{rastqic14}.
Measured results of the experiment with fixed  $\theta$ do
violate the inequality $\tca_{q}\leq0$ for one values of $q$ and
do not for other ones, including the standard case $q=1$. It is
a manifestation of the following fact. Entropic Leggett--Garg
inequalities give only necessary conditions that probabilistic
models are compatible with the macrorealistic picture.

For larger values of $s$ or $n$, a similar situation is observed.
Here, we refrain from presenting corresponding curves. Instead, we
describe some significant points. The above mentioned properties
of curves for different $q$ remain valid. In particular, there is
some domain, in which $q$-entropic inequalities give advances in
comparison with the standard case $q=1$. On the other hand, with
growing $s$ and $n$ we have seen a decrease of this domain. It may
be related with the following fact. As reported in Ref.
\cite{uksr12}, both the strength and the range of violations
reduce with the increase of spin value. We also recall that the
considered situation corresponds to equidistant time intervals.
For experiments with unequal time intervals, the $q$-entropic
approach may give additional possibilities for analyzing data of
tests of the Leggett--Garg type. Another question is related to
detection inefficiencies.

Using the Shannon entropies, the writers of Ref. \cite{rchtf12}
examined the Bell inequalities in the case of detection
inefficiencies. For the $q$-entropic inequalities, this issue was
studied in Ref. \cite{rastqic14}. We showed that the $q$-entropic
approach can allow to reduce an amount of required detection
efficiency. We shall now examine this question for restrictions
of the Leggett--Garg type. In the considered example, we have the
probability (\ref{prmj}), whence
\begin{equation}
H_{q}(X_{k})=\ln_{q}(2s+1)
\ . \label{hq2s}
\end{equation}
Then the additional term (\ref{etac}) reads
\begin{align}
\Delta_{q}(\eta)&=(n-2)
\Bigl\{
\eta^{q}\bigl(\eta^{q}+2(1-\eta)^{q}-1\bigr)\ln_{q}(2s+1)
\Bigr.
\nonumber\\
\Bigl.
&+\bigl(\eta^{q}+(1-\eta)^{q}\bigr)h_{q}(\eta)\Bigr\}
\ . \label{etac2}
\end{align}
The characteristic quantity $\ca_{q}$ is given by the numerator of
Eq. (\ref{tcdf}). The $q$-entropic inequality (\ref{cmb12}) claims
$\ca_{q}\leq0$. Using measured data, we will actually deal with
the quantity (\ref{caet}). As was mentioned above, we must confide
that the violating term $\eta^{2q}{\,}\ca_{q}$ is sufficiently
large in comparison with the additional term (\ref{etac2}). To do
so, we introduce their ratio
\begin{equation}
r_{q}(\eta):=\eta^{-2q}{\,}\ca_{q}^{-1}\bigl|\Delta_{q}(\eta)\bigr|
\ , \label{rat}
\end{equation}
which is restricted to the case $\ca_{q}>0$. Let us consider this
ratio in our case $s=1/2$ and $n=3$. We use $\theta=0.9$, when the
strength of violations is large for several values of $q$ (see
Fig. \ref{fig1}). We have calculated $r_{q}(\eta)$ versus $\eta$
for such values of $q$. With respect to $\eta$, we especially
focus an attention on values, which are close to $1$ from below.
For fixed $q$, the ratio $r_{q}(\eta)$ decreases with such $\eta$
almost linearly, up to the inefficiency-free value $r_{q}(1)=0$.
Due to almost linear dependence, we can describe each case by the
value (\ref{rat}) for some $\eta$, say, for $\eta=0.99$.
Approximately, we use $r_{q}(\eta)\approx
10^{2}{\,}r_{q}(0.99){\,}(1-\eta)$ within a range of linear
behavior. In Table \ref{tab1}, the value $r_{q}(0.99)$ is
presented for $\theta=0.9$ and several values of $q$.

\begin{table}
\begin{center}
\caption{\label{tab1}The values of the ratio (\ref{rat}) for
$\eta=0.99$ and several $q$ in the case $s=1/2$, $n=3$, and $\theta=0.9$.}
\vskip0.1cm
\begin{tabular}{|c|c|c|c|c|c|c|c|c|c|c|c|c|c|}
\hline
$q$ & 1.0 & 1.1 & 1.2 & 1.4 & 1.6 & 1.8 & 2.0  \\
\hline
$r_{q}(0.99)$ & 0.711 & 0.504 & 0.386 & 0.266 & 0.212 & 0.186 & 0.173 \\
\hline
\hline
$q$ & 2.2 & 2.4 & 3.0 & 4.0 & 6.0 & 8.0 & 10.0 \\
\hline
$r_{q}(0.99)$ & 0.167 & 0.165 & 0.174 & 0.208 & 0.330 & 0.587 & 1.364 \\
\hline
\end{tabular}
\end{center}
\end{table}

In general, the required amount of efficiency seems to be very
high. At the same time, the value $r_{q}(0.99)$ essentially
depends on $q$. Initially, this value quickly decreases with
$q>1$. It then becomes increasing for sufficiently large $q$.
Among $q$-entropic inequalities of the Leggett--Garg type, the
choice $q=2$ is convenient. First, both the strength and range of
violations are significant. Second, the ratio (\ref{rat}) is small
for $\eta>0.99$ (see Table \ref{tab1}). Third, properties of the
$q$-entropies are mathematically simpler in the case $q=2$. We
have already reported such reasons in Ref. \cite{rastqic14}, where
$q$-entropic inequalities of the Bell type were obtained. Studying
the CHSH and KCBS scenarios, the $q$-entropic Bell inequalities
were shown to be expedient. Using $q$-entropic inequalities, we
can also reach new possibilities for analyzing data of the
Leggett--Garg tests. Leggett--Garg inequalities under decoherence
also deserve theoretical studies. We hope to address this question
in future investigations.

\section{Conclusions}\label{sec5}

We have formulated inequalities of the Leggett--Garg type in terms
of the $q$-entropies. For all $q\geq1$, such inequalities follow
from the existence of some joint probability distribution for
outcomes of measurements at different instants of time. It turned
out that quantum mechanics predicts violations of an entire family
of $q$-entropic inequalities of the Leggett--Garg type. We
illustrated violations with the example of quantum spin systems.
The spin-$s$ system has been prepared initially in the completely
mixed state. Entropic Leggett--Garg inequalities give only
necessary conditions that probabilistic models are compatible with
the macrorealism in the broader sense. We showed that the
presented inequalities allow to widen a class of situations, in
which an incompatibility with the macrorealism can be checked.
Both the strength and range of violations can be increased by
adopting appropriate values of the entropic parameter $q$. We also
formulated $q$-entropic inequalities of the Leggett--Garg type in
the case of detection inefficiencies. If we use $q$-entropic
inequalities, then the required amount of efficiency can somehow
be reduced. In the sense of experimental testing, Leggett--Garg
inequalities for quantum system in a dephasing environment are
another important question \cite{pal10,xu11}. The presented
formulation could be useful in analysis of recent experiments to
test Leggett--Garg inequalities.

\end{document}